\documentclass[noshowkeys,floatfix,twocolumn,nofootinbib,aps,prl,10pt, superscriptaddress,longbibliography]{revtex4-2}
\usepackage[utf8]{inputenc} 
\usepackage[english]{babel} 
\usepackage[normalem]{ulem} 
\usepackage{amsfonts} 
\usepackage{amsmath} 
\usepackage{amssymb} 
\usepackage{bbm} 
\usepackage{physics} 
\usepackage{braket} 
\usepackage{bm} 
\usepackage{cases} 
\usepackage{tikz}
\usetikzlibrary{positioning,decorations.pathmorphing, decorations.pathreplacing, decorations.shapes}
\usepackage[compat=1.1.0]{tikz-feynhand}
\usepackage{calc}
\usepackage{graphicx} 
\usepackage{xcolor} 
\usepackage[caption=false,position=top]{subfig}
\usepackage{xr}
\usepackage[citecolor=blue, linkcolor=blue, urlcolor=blue, filecolor=blue, hypertexnames=true, colorlinks, bookmarksopenlevel=1, pdfstartview=FitH, bookmarksopen, bookmarksnumbered, plainpages=false, linktocpage]{hyperref}
\let\oldciteauthor=\citeauthor
\def\citeauthor#1{{\hypersetup{citecolor=black}\oldciteauthor{#1}}}
\let\oldfootnote=\footnote
\def\footnote#1{\hypersetup{citecolor=blue}\oldfootnote{#1}}
\usepackage[capitalize]{cleveref}
\usepackage{url}

\def\cal#1{\mathcal{#1}}

\newcommand{\opc}{\hat{c}}

\newcommand{\T}[1]{\textrm{#1}}

\newcommand{\br}{{\bm r}}

\newcommand{\bk}{\bm{k}}
\newcommand{\bq}{\bm{q}}
\newcommand{\bpos}{\bm{r}}
\newcommand{\bG}{\bm{G}}

\newcommand{\Qm}{\mathcal{Q}}

\newcommand{\Fm}{\mathcal{F}}

\tikzset{
    between/.style args={#1 and #2}{
         at = ($(#1)!0.5!(#2)$)
    }
}
\makeatletter
\pgfdeclaredecoration{bettercoil}{coil}
{
  \state{coil}[switch if less than=%
    0.5\pgfdecorationsegmentlength+
    \pgfdecorationsegmentaspect\pgfdecorationsegmentamplitude+%
    \pgfdecorationsegmentaspect\pgfdecorationsegmentamplitude to last,
               width=+\pgfdecorationsegmentlength]
  {
    \pgfpathcurveto
    {\pgfpoint@oncoil{0    }{ 0.555}{1}}
    {\pgfpoint@oncoil{0.445}{ 1    }{2}}
    {\pgfpoint@oncoil{1    }{ 1    }{3}}
    \pgfpathcurveto
    {\pgfpoint@oncoil{1.555}{ 1    }{4}}
    {\pgfpoint@oncoil{2    }{ 0.555}{5}}
    {\pgfpoint@oncoil{2    }{ 0    }{6}}
    \pgfpathcurveto
    {\pgfpoint@oncoil{2    }{-0.555}{7}}
    {\pgfpoint@oncoil{1.555}{-1    }{8}}
    {\pgfpoint@oncoil{1    }{-1    }{9}}
    \pgfpathcurveto
    {\pgfpoint@oncoil{0.445}{-1    }{10}}
    {\pgfpoint@oncoil{0    }{-0.555}{11}}
    {\pgfpoint@oncoil{0    }{ 0    }{12}}
  }
  \state{last}[next state=final]
  {
    \pgfpathcurveto
    {\pgfpoint@oncoil{0    }{ 0.555}{1}}
    {\pgfpoint@oncoil{0.445}{ 1    }{2}}
    {\pgfpoint@oncoil{1    }{ 1    }{3}}
    \pgfpathcurveto
    {\pgfpoint@oncoil{1.555}{ 1    }{4}}
    {\pgfpoint@oncoil{2    }{ 0.555}{5}}
    {\pgfpoint@oncoil{2    }{ 0    }{6}}
  }
  \state{final}{}
}

\def\pgfpoint@oncoil#1#2#3{%
  \pgf@x=#1\pgfdecorationsegmentamplitude%
  \pgf@x=\pgfdecorationsegmentaspect\pgf@x%
  \pgf@y=#2\pgfdecorationsegmentamplitude%
  \pgf@xa=0.083333333333\pgfdecorationsegmentlength%
  \advance\pgf@x by#3\pgf@xa%
}
\makeatother

\begin{document}


\date{\today}
\title{A universal route to chiral  Ising superconductivity in monolayer   TaS\texorpdfstring{$_2$}{2} and NbSe\texorpdfstring{$_2$}{2}}
\author{Lucia Gibelli}
\affiliation{Institute for Theoretical Physics, University of Regensburg, 93 053 Regensburg, Germany}
\author{Simon Höcherl}
\affiliation{Institute for Theoretical Physics, University of Regensburg, 93 053 Regensburg, Germany}
\author{Julian Siegl}
\affiliation{Institute for Theoretical Physics, University of Regensburg, 93 053 Regensburg, Germany}
\author{Viliam  Va\v{n}o}
\affiliation{Department of Applied Physics, Aalto University, FI-00076 Aalto, Finland}
\author{Somesh C.~Ganguli }
\affiliation{Department of Applied Physics, Aalto University, FI-00076 Aalto, Finland}
\author{Magdalena Marganska} 
\affiliation{Department of Theoretical  Physics, Wrocław University of Science and Technology, Wybrzeże Wyspiańskiego 27, 50-370 Wrocław, Poland}
\author{Milena Grifoni}
\email{Milena.Grifoni@ur.de}
\affiliation{Institute for Theoretical Physics, University of Regensburg, 93 053 Regensburg, Germany}
\begin{abstract}
We investigate Ising superconductivity in two archetypal intrinsic  superconductors, 
monolayer 1H-TaS$_2$ and 1H-NbSe$_2$, in a bottom-up approach. 
Using ab initio-based tight-binding parameterizations for the relevant low-energy $d$-bands, the screened interaction is evaluated microscopically, in a scheme including Bloch overlaps. 
In direct space, the screened potential displays for both systems long-range Friedel oscillations alternating in sign. Upon scaling, the oscillation pattern becomes universal, with the periodic features locked to the lattice.  
 Solving the momentum-resolved gap equations, a chiral ground state with $p$-like symmetry is generically found. 
Due to the larger Ising spin-orbit coupling,
the chiral gap is more anisotropic in  
TaS$_2$ than in NbSe$_2$. This is reflected in tunneling spectra displaying V-shaped features for the former, in quantitative agreement with low-temperature  scanning tunneling experiments on TaS$_2$.  At the same time, our results reconcile the apparent discordance with hard gap tunneling spectra reported for the sibling NbSe$_2$.    
\end{abstract}
\maketitle

\textit{Introduction}---Few-layer transition metal dichalcogenides (TMDCs) offer unprecedented opportunities to design novel quantum materials by gating and  proximity effects with substrates, impurity doping or by exploiting their strong spin-orbit coupling (SOC). This also concerns their superconducting properties~\cite{zhangIsingPairingAtomically2021}. 

\begin{figure}[ht]
   \includegraphics[width=\columnwidth]{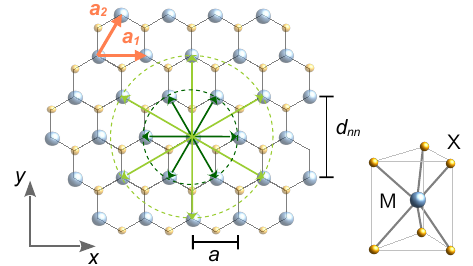}
    \caption{
        {{Crystalline structure of monolayer 1H-TMDCs seen from the top}} with the metal (blue) and the chalcogen (yellow) atoms. The electronic structure at the Fermi level is mostly determined by the $d$-orbitals of the metal atom.  The latter atoms  span a triangular lattice with  generators $\bm{a}_1$ and $\bm{a}_2$. The distance between neighboring metal atoms equals the lattice spacing $a$ in the $x$-direction while is $d_{nn}=\sqrt{3}a$ in the $y$-direction.  
    } 
    \label{fig:lattice}
\end{figure}

Among the most investigated intrinsic TMDCs superconductors are NbSe$_2$ and TaS$_2$, which display charge-density waves  already in their bulk  phase~\cite{wilsonChargedensityWavesSuperlattices1975a,nagataSuperconductivityLayeredCompound1992a,sannaRealspaceAnisotropySuperconducting2022}
 and  show  signatures of unconventional superconductivity  in the few-layer limit~\cite{xiIsingPairingSuperconducting2016,xingIsingSuperconductivityQuantum2017,navarro-moratallaEnhancedSuperconductivityAtomically2016,delabarreraTuningIsingSuperconductivity2018}. 
The diverse properties of TMDCs are governed
by their crystal phases~\cite{chhowallaChemistryTwodimensionalLayered2013a}. 
Here we focus on the trigonal prismatic 1H-TMDCs. 
In their bulk form, these TMDCs possess a global inversion center, but are non-centrosymmetric at the monolayer level. Viewed from above, the monolayer crystal forms a hexagonal lattice similar to graphene but with two inequivalent sublattice sites occupied by the M and X atoms, see Fig.~\ref{fig:lattice}.  The lack of inversion symmetry 
leads  to  Ising SOC, pinning the electron spins in the out-of-plane direction~\cite{xuSpinPseudospinsLayered2014,xiIsingPairingSuperconducting2016,zhangIsingPairingAtomically2021}, which qualitatively impacts key features of the superconducting state. For example, it leads to a violation of the Pauli limit under strong in-plane magnetic fields~\cite{zhangIsingPairingAtomically2021,frigeriSuperconductivityInversionSymmetry2004,xiGateTuningElectronic2016,ilicEnhancementUpperCritical2017,delabarreraTuningIsingSuperconductivity2018,mockliMagneticfieldInduced$s+mathitif$2019,choNodalNematicSuperconducting2022,engstromUpperCriticalField2025}, 
and yields the potential to realize exotic topological phases \cite{heMagneticFieldDriven2018,chenTopologicalIsingPairing2019,shafferCrystallineNodalTopological2020}. 

Despite the last years having witnessed a surge of growing interest in Ising superconductivity of monolayer TMDCs, the pairing mechanisms are still actively discussed. They range from phonon-mediated ~\cite{hinscheElectronPhononInteraction2017,zhengElectronphononCouplingCoexistence2019,dasElectronphononCouplingSpin2023} to more exotic ones resulting from competing electronic repulsion, where even predictions on the symmetry of the gap may differ~\cite{wickramaratneIsingSuperconductivityMagnetism2020,shafferCrystallineNodalTopological2020,horholdTwobandsIsingSuperconductivity2023,royUnconventionalPairingIsing2024,sieglFriedelOscillationsChiral2025}. 
When electron pairing arises from virtual phonon exchange~\cite{bardeenTheorySuperconductivity1957}, 
the  ions mediate an effective attractive interaction 
that favors $s$-wave pairing, 
 with gap functions isotropic in momentum space.
If superconductivity emerges from screened repulsion, the pairing function is anisotropic~\cite{scalapinoCommonThreadPairing2012}, 
 and its symmetry is associated to irreducible representations of the crystal. 
Two-dimensional (2D) systems with triangular lattices allow two-dimensional representations, and  are potential candidates for becoming chiral superconductors~\cite{mackenzieEvenOdderTwentythree2017,black-schafferChiralDwaveSuperconductivity2014,nandkishoreChiralSuperconductivityRepulsive2012}. 

 This uncertainty on the nature of superconductivity in
few-layer TMDCs is partly due to the inherent difficulty
of a realistic modeling of many-body effects, which properly accounts for the reduced  screening of the Coulomb interaction in low dimensions ~\cite{hybertsenElectronCorrelationSemiconductors1986,rohlfingElectronholeExcitationsOptical2000}. Latest \textit{ab-initio} studies have confirmed the long-range nature of the screened interaction 
in monolayer TMDCs~\cite{ramezaniNonconventionalScreeningCoulomb2024}.

In a recent work~\cite{sieglFriedelOscillationsChiral2025}, some of us have proposed a {\it{bottom-up}} approach to unconventional superconductivity in monolayer NbSe$_2$ involving a microscopic  calculation of the screened Coulomb interaction. Within a generalized random phase approximation (RPA) in momentum space, a strong suppression of intravalley over intervalley processes due to the   electron-hole fluctuations of the metal was demonstrated. In the real space, this gives rise to 
 long-range Friedel oscillations~\cite{friedelXIVDistributionElectrons1952} of the screened Coulomb potential alternating in sign.
Cooper pairing among the electrons then emerges when taking advantage of the attractive regions, a manifestation of the 
 pure electronic mechanism for superconductivity proposed by Kohn and Luttinger ~\cite{kohnNewMechanismSuperconductivity1965}. 
By solving self-consistent gap equations, the appearance of a chiral ground state with $p$-wave character was found.

Triggered by the above results, some fundamental questions arise which will be investigated in this Letter.
The first one concerns the nature of superconductivity in monolayer 1H-TaS$_2$, another archetypal TMDC superconductor. Since the band structure is similar, one can still expect the  unconventional Kohn-Luttinger mechanism to apply as for the NbSe$_2$. However, will the stronger Ising SOC~\cite{delabarreraTuningIsingSuperconductivity2018} also modify the nature of the ground state? If the gap is still chiral, how can one reconcile this with the seemingly different scanning tunneling spectroscopy (STS)  spectra reported for the two systems? While  low-temperature STS data for 1H-NbSe$_2$ are compatible with a fully gapped 
$s$-like or  chiral phase~\cite{wanObservationSuperconductingCollective2022},  the STS spectra for 1H-TaS$_2$ have rather a V-like shape \cite{vanoEvidenceNodalSuperconductivity2023}.

 In this work we have applied  the methodology from~\cite{sieglFriedelOscillationsChiral2025} to evaluate the screened interaction in 1H-TaS$_2$. To identify universal features, for both test systems the investigation of the real-space interaction was extended to several lattice periods.  
 We found that, upon proper scaling, the Friedel oscillations of NbSe$_2$ and TaS$_2$ display almost identical behavior, with anisotropic oscillations   dictated by the $D_{\rm 3h}$ symmetry of the underlying crystalline lattice.   In a 2D Fermi liquid, the Kohn-Luttinger anomaly generically leads to $p$-wave pairing~\cite{chubukovKohnLuttingerEffectInstability1993}. By solving the gap equations, we confirm the emergence of a $p$-like chiral ground state also for 1H-TaS$_2$.  Importantly, the larger Ising SOC arising from the Ta atoms yields a stronger anisotropy of the chiral gap  compared to  1H-NbSe$_2$. This  results in  V-shaped tunneling spectra for TaS$_2$, despite the gapped nature of the chiral phase, similar to the STS experiments~\cite{vanoEvidenceNodalSuperconductivity2023}. 
Our results establish the Kohn-Luttinger  mechanism as possible route to superconductivity in 2D TMDCs, and  identify in the distinct strength of the Ising SOC the reason for  seemingly different low-temperature STS spectra.

\begin{figure}[t]
\includegraphics[width=\columnwidth]{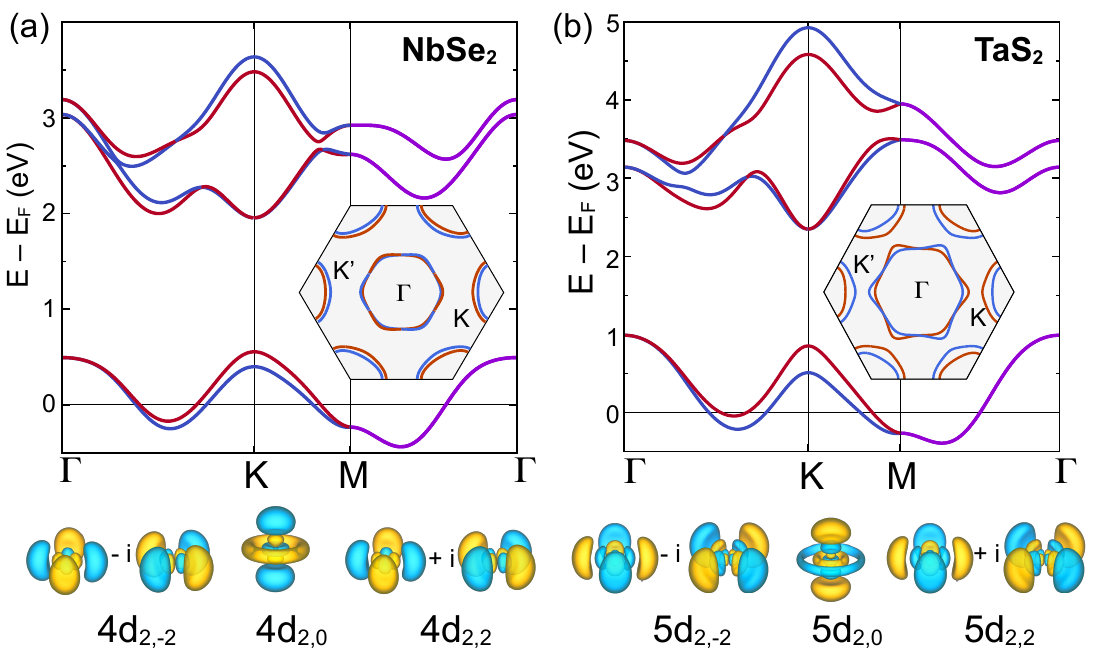}
 \caption{
    {Band structure, Fermi surfaces and orbital contributions in monolayer
    1H-NbSe$_{2}$ and 1H-TaS$_{2}$. (a), (b)}
    Tight-binding Bloch bands (top) and Fermi surfaces (inset) with 
      spin degeneracy removed by the Ising spin-orbit coupling. 
  Due to the larger spin-orbit coupling, the $\Gamma$- and $K$-Fermi surfaces get closer in TaS$_2$ compared to NbSe$_2$.   (bottom) Sketch of the three atomic $d_{l,m}$ orbitals used in the calculation. 
    The quantum numbers $l,m$ denote the total angular momentum of the orbital and its \hbox{azimuthal} projection.
    \textcolor{red}{}}
    \label{fig:bands}
\end{figure}

\noindent {\textit{Band structure and orbital composition}---}We start from  a tight-binding model for the Bloch band structure of metallic 1H-TMDCs,  like NbSe\texorpdfstring{$_2$}{2} or TaS\texorpdfstring{$_2$}{2}, obtained by using their three most relevant $d$-orbitals~\cite{liuThreebandTightbindingModel2013} fitted to ab-initio calculations~\cite{heMagneticFieldDriven2018,margalitTheoryMultiorbitalTopological2021}, cf.  End Matter for details. 
As seen in Fig.~\ref{fig:bands}, 
there are two metallic  bands relevant for superconducting pairing at the Fermi level which are split by Ising SOC~\cite{liuThreebandTightbindingModel2013}.
Multiple disjoint Fermi surfaces are present, shown in the insets of Fig. \ref{fig:bands}(a), (b). 
While the planar $d$-orbitals mostly contribute to the band composition around the $K$ and $K$' valleys, $d_{2,0}$ is dominant at the $\Gamma$ valley~\cite{liuThreebandTightbindingModel2013}. The impact of the Ising SOC is clearly more prominent in TaS$_2$ than in NbSe$_2$. 

We restrict to the spin-split metallic bands at the Fermi level,  for which the many-body Hamiltonian in the Bloch basis has the form  
\begin{align}
    \label{singleband interacting Hamiltonian}
    \hat{H}_{\rm MB}=\sum_{\bk,\sigma} \xi_{\bk,\sigma}\opc_{\bk,\sigma}^\dagger \opc_{\bk,\sigma}+\hat{V}\,, 
\end{align}
where the first term describes independent Bloch electrons  with energies $\xi_{\bk,\sigma}$ measured from the Fermi level, and 
$\opc_{\bk,\sigma}^{(\dagger)}$ 
are  annihilation (creation)  operators of Bloch electrons with crystal momentum $\bk$ and spin $\sigma$. 
The second term is the screened Coulomb interaction. Its form is crucial  to assess whether a purely electronic pairing mechanism can take place.  
%
%
\begin{figure*}[t]
   \includegraphics[width=\textwidth]{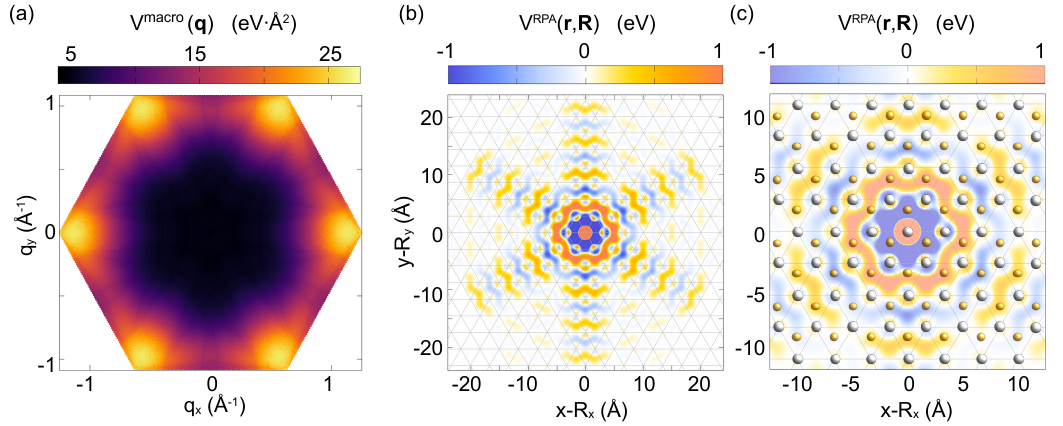}
    \caption{
        {{Screened RPA interaction for TaS$_2$.}} 
        (a) In the reciprocal space the screened interaction is positive and inherits the $D_{3\rm h}$ symmetry of the lattice.  Exemplary the element of the interaction matrix corresponding to normal processes  is shown.  It is strongly suppressed at small transferred momenta $\bq_\Gamma$ while  maxima are seen for  momenta $\bq_K$ at the corners of the Brillouin zone. (b) In the real space the interaction is largely anisoptropic and displays Friedel oscillations with alternating positive and negative regions. The oscillation persist at longest (shortest) along the $y$ $(x)$-axis (and the lines obtained from those by a $2\pi/6$  rotation). A zoom into the central area of (b) is shown in (c), overlaid with the Ta and S atomic lattice. 
    } 
    \label{fig: Friedel_Ta} 
\end{figure*}

\begin{figure}[t]
   \includegraphics[width=\columnwidth,height=9cm]{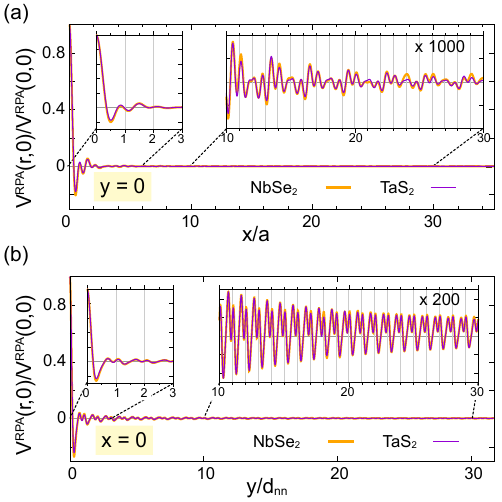}
    \caption{
        {{Universality of Friedel oscillations.}}
        The screened potentials for  TaS$_2$ and NbSe$_2$ almost collapse onto the same curve when scaled by their value at the origin and plotted in units of the respective lattice constant.  
 The oscillations persist over long distances and have a periodicity  $3a$ in the $x$ direction and  $d_{nn}=\sqrt{3}a$ in the $y$ direction. The lattice spacing are taken to be $a_{\rm{TaS}_2}=3.342~\rm{\AA}$, $a_{\rm{NbSe}_2}=3.445~\rm{\AA}$, the interactions  $V^{\rm{RPA}}_{\rm{TaS}_2}(0,0)=24.33$~eV,  $V^{\rm{RPA}}_{\rm{NbSe}_2}(0,0)=23.07$~eV.  } 
    \label{fig: oscillations comparison_TaNb} 
\end{figure}

\noindent{\textit{Screened Coulomb interaction and Friedel oscillations}}---In crystalline systems, the screened interaction $V(\bm{r},\bm{r}')$ is in  general a function of both positions of the two interacting charges. 
A closed form evaluation of the screened interaction  can be obtained in the RPA, in which the polarizability is approximated by that of independent Bloch electrons~\cite{hybertsenElectronCorrelationSemiconductors1986}.
In monolayer TMDCs, with the $d$-orbitals tightly confined to the plane of the material, one finds the expansion~\cite{sieglFriedelOscillationsChiral2025} 
\begin{align}
    &V^{\rm RPA}(\bm{r},\bm{r}')
    &=\frac{1}{\cal{A}}\sum_{\bq,\bm{G},\bm{G}'}V^{2{\rm D}}_{\bm{G},\bm{G}'}(\bq)e^{i(\bq+\bm{G})\cdot\bpos}e^{-i(\bq+\bm{G}')\cdot\bpos'}\,,
\end{align}
where $\bG$, $\bm{G}'$ are reciprocal lattice vectors, $\bq$ is a crystal momentum,  $\cal{A}$ is the sample area, and $ V_{\bm{G},\bm{G}'}^{\T{2D}}(\bm{q})$ is the screened interaction tensor. 
 In Fig.~\ref{fig: Friedel_Ta}, we display the screened RPA interaction for  TaS$_2$.   The element $V^{\rm{macro}}(\bm{q}):=V_{\bm{0},\bm{0}}^{\T{2D}}(\bm{q})$   
is shown in  Fig.~\ref{fig: Friedel_Ta}(a), and  provides the macroscopic, long-range component of the interaction.  
Similarly to what was reported  for NbSe$_2$ in~\cite{sieglFriedelOscillationsChiral2025},   it is strongly suppressed for $\bq\approx\bq_\Gamma$, and displays maxima at $\pm\bq_K$, which 
 supports the scenario of unconventional pairing arising from dominant intervalley scattering \cite{roldanInteractionsSuperconductivityHeavily2013,horholdTwobandsIsingSuperconductivity2023}.
When looking at the potential in the real space, $V^{\rm RPA}(\br,\bm{R})$, with $\br'=\bm{R}$ fixed to a Ta site, alternating repulsive and attractive regions are seen, cf. Fig.~\ref{fig: Friedel_Ta}(b).  %
A large anisotropy is observed which respects the $D_{3 \rm h}$ symmetry of the lattice with  very different decay  along preferential directions. 
One finds on-site repulsion at the origin and minima or maxima  about halfway to neighboring atom sites, Fig.~\ref{fig: Friedel_Ta}(c). Hence, the  Friedel oscillations are {\it{locked to the lattice}}, 
in contrast to the case of the interacting electron gas studied by Kohn and Luttinger for simple metals~\cite{kohnNewMechanismSuperconductivity1965}.  We find a very similar oscillation pattern when calculating the screened interaction for NbSe$_2$, which hints at a universal behavior upon proper scaling. This is confirmed in Fig.~\ref{fig: oscillations comparison_TaNb}, where the screened interactions for NbSe$_2$ and TaS$_2$  almost collapse onto the same curve when divided by their value at the origin and plotted in units of the lattice constant. The oscillations persist  at very long distances,  whereby the longer lasting ones along the $y$-axis are periodic with the next neighbor distance $d_{nn}=\sqrt{3}a$, and the faster decaying ones along the $x$-axis have the period $3a$. 
This general behavior suggests that the two materials should share similar features also in the superconducting state.

{\textit{Gap equation and leading instabilities near $T_{\rm c}$}}---We next discuss the properties of the superconducting instability mediated by the screened potential. With this aim, we conveniently work in the Bloch basis. Time-reversal symmetry favors scattering between time-reversal partners, as shown in the End Matter. 
We hence consider in the following the interaction matrix elements 
${V_{\bk,\bk',\sigma}:=\bra{\bk',\sigma';\bar{\bk}',\bar\sigma'}{\hat{V}}\ket{\bk,\sigma;\bar{\bk},\bar{\sigma}}}$ where $\bk,\bk'$  are in-plane momenta within the first Brillouin zone.
The restriction to Kramer pairs of Bloch states yields a low-energy Hamiltonian of the typical BCS form~\cite{bardeenTheorySuperconductivity1957}.
Performing a mean field approximation, in turn, leads to the familiar BCS gap equation 
\begin{align}
    \label{gap equation trs}
    &\Delta_{\bk,\sigma}=-\sum_{\bk'}V_{\bk,\bk',\sigma}\Pi(E_{\bk',\sigma})\Delta_{\bk',\sigma}\,,
\end{align}
where the  gap enters $\Pi(E)=\tanh(\beta E/2)/2E$  through the quasiparticle energies
$E_{\bk,\sigma}=\sqrt{\xi_{\bk,\sigma}^2+\vert\Delta_{\bk,\sigma}\vert^2}$, and $\beta=1/k_BT$ is the inverse temperature. 

\begin{figure}[t]
   \includegraphics[width=\columnwidth]{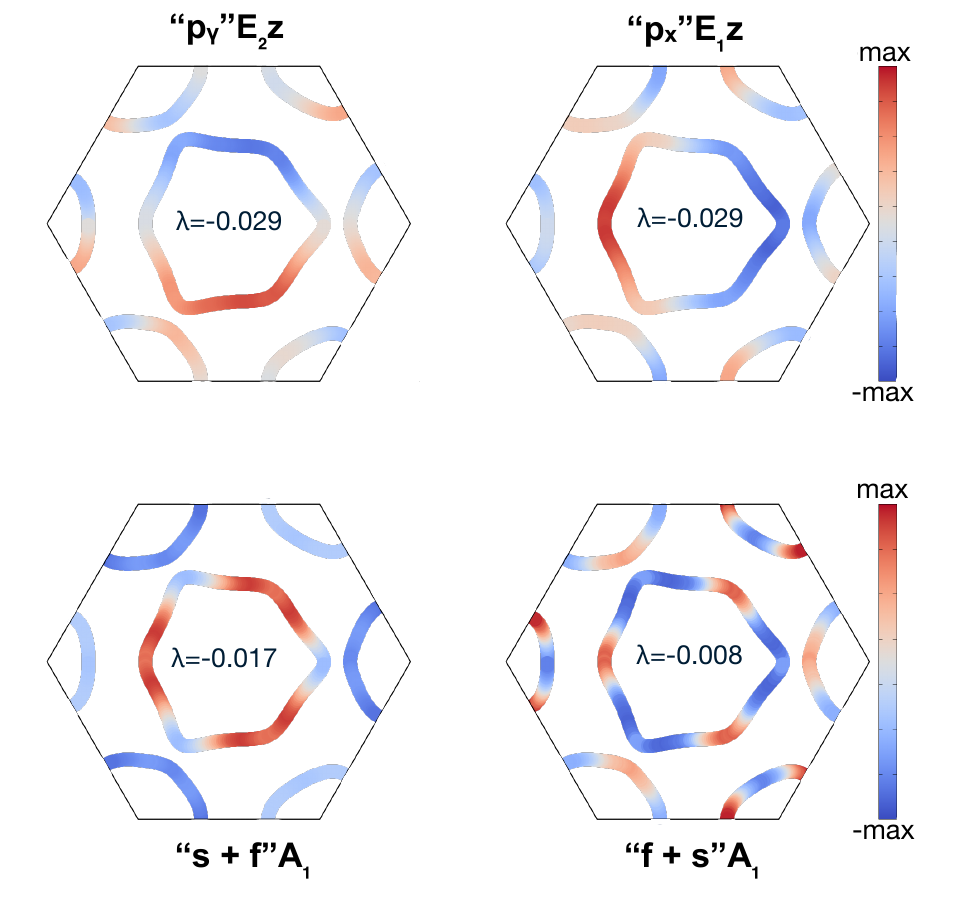}
    \caption{
        {{Instabilities near $T_{\rm c}$ for TaS$_2$.}}
        At the critical temperature the leading gaps are two-fold degenerate solutions of predominant $E'$ $z$-triplet type (top panels). They are followed by SOC-allowed linear combinations  of the $A_1'$ type (bottom panels). Due to the strong admixture of $s$-like and $f$-like contributions, these gaps have nodes on the   $\Gamma$ Fermi surfaces. A complex linear combination of the nematic $E'$ solutions results in a fully gapped chiral gap.
    } 
    \label{fig: Instabilities_Ta} 
\end{figure}

To access observables near the phase transition, one linearizes~\cref{gap equation trs} by replacing the excitation energies $E_{\bk,\sigma}$ with the single-particle energies $\xi_{\bk,\sigma}$ and solving the resulting linear problem in a range of $\pm \Lambda$  around the Fermi energy~\cite{shafferCrystallineNodalTopological2020,royUnconventionalPairingIsing2024}. 
The instabilities with the highest $T_{\T{c}}$ for NbSe$_2$ 
have been  investigated  in~\cite{royUnconventionalPairingIsing2024} using  a top-down Hubbard-Kanamori model in the atomic basis, and in~\cite{sieglFriedelOscillationsChiral2025} within the bottom-up approach also adopted in this work. Here we focus  on  the leading instabilities of 1H-TaS$_2$, which 
are shown in \cref{fig: Instabilities_Ta}. 
They belong to irreducible representations of the $D_{\mathrm{3h}}$ symmetry group. 
We find as the dominant pairing   two degenerate nematic $p$-like solutions of the $E'$ $z$-triplet type (with singlet admixture). Similar conclusions have just been reported in a top-down model calculation by Roy {\it{et al.}}~\cite{royUnconventionalSuperconductivityMonolayer2025}, which has appeared while submitting this work. 
In our work these solutions are followed by the SOC-introduced combinations of the $A'_1$ $s$-like singlet and $f$-like $z$-triplet, with the latter a previously discussed candidate for the superconducting phase of monolayer TaS$_2$~\cite{vanoEvidenceNodalSuperconductivity2023}.
The critical temperatures of these solutions depend on the associated eigenvalues $\lambda$ of the pairing matrix and the cutoff in energy 
which we fixed to $\Lambda=100\,\T{meV}$. 
 
{\it{Chiral $p_x+ip_y$ groundstate and comparison with experiments}}--- The two leading solutions of the pristine system are degenerate nematic gaps, which will remain   quasi-degenerate for weak enough breaking of the symmetry. 
This allows for the emergence of a chiral gap, i.e., a complex linear combination of the two nematic solutions,  at temperatures below $T_{\T{c}}$~\cite{kallinChiralSuperconductors2016}. 
 \begin{figure}[t]
   \includegraphics[width=\columnwidth]{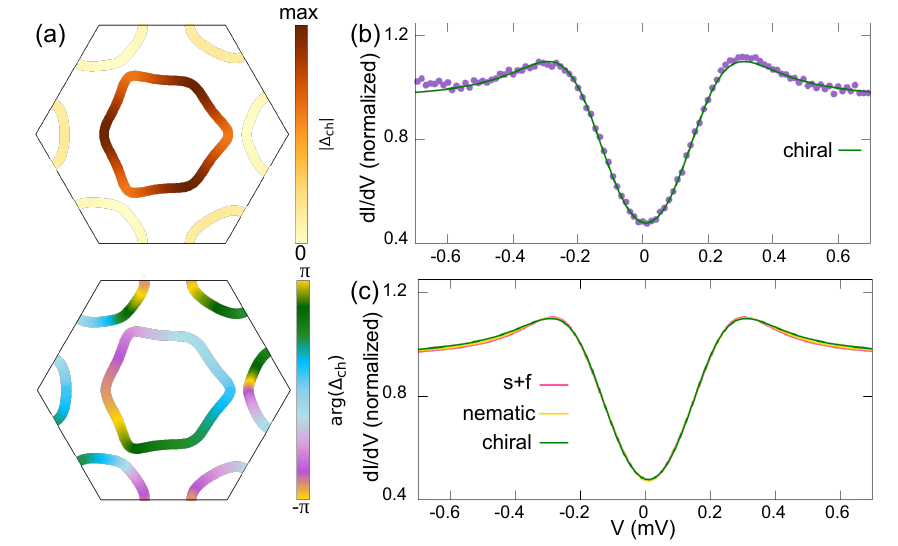}
    \caption{
        {{Chiral gap and STS spectra for TaS$_2$. }} (a) 
   A complex linear combination of the two  $E'$ $z$-triplets provides the groundstate of the system. It is fully gapped  with a phase winding clockwise around the $\Gamma$ Fermi surface. (b), (c)  Experimental spectra recorded at $T=350$~mK are well fitted with a theoretically evaluated differential conductance which uses the chiral gap. A similar good agreement is found if the nematic or the $s+f$ solutions are used since the associated best fit theoretical curves are very close. 
    } 
    \label{fig: STS_Ta} 
\end{figure} 
To what extent a nodal nematic solution or the gapped chiral phase is the preferred ground state, it can be determined by  solving the self-consistency equation, \cref{gap equation trs}, as a function of temperature.
 To this aim we expanded both the interaction and the order parameter in basis functions of the appropriate irreducible representations of the $D_{3\rm{h}}$ symmetry group~\cite{shafferCrystallineNodalTopological2020,shafferWeakcouplingTheoryPair2023,hanisDistinguishingNodalNonunitary2024}, and solved a set of coupled equations for the expansion coefficients as a function of the temperature.
We find a fully gapped chiral phase at zero temperature, similarly to what was 
predicted for monolayer NbSe$_2$~\cite{sieglFriedelOscillationsChiral2025}. 
The magnitude of this chiral gap at $T=0\,\T{mK}$ is displayed in Fig.~\ref{fig: STS_Ta} together with a comparison with the experimental STS data from~\cite{vanoEvidenceNodalSuperconductivity2023}. 
As seen in Fig.~\ref{fig: STS_Ta}(a), the chiral gap is strongly dispersive on the $\Gamma$ Fermi surface, where it also attains its maximal value. 
The experiments were performed on monolayer 1H-TaS$_2$ epitaxially grown on a highly-oriented pyrolitic graphite (HOPG) substrate, see~\cite{vanoEvidenceNodalSuperconductivity2023} for details of the setup. 
The measured observable is the  differential tunneling conductance ($\T{d}I/\T{d}V$), which is proportional to the spectral function of the sample. 
The experimental $\T{d}I/\T{d}V$  averaged over 12 spectra measured at different locations at $350\,\T{mK}$ is shown in Fig.~\ref{fig: STS_Ta}(b) and fitted to the tunneling density of states (DOS) obtained from the chiral gap.
The theoretical differential conductance shown in Fig.~\ref{fig: STS_Ta}(b) is calculated using the coupling to the $\Gamma$ and $K$ Fermi pockets as free parameters. A table of  the used parameters is found in  the End Matter. We find  a much larger coupling to the states on the $\Gamma$ Fermi surface compared to those near the $K$ and $K'$ valleys, in agreement with the sensitivity of STM to states with small in-plane momentum~\cite{tersoffTheoryApplicationScanning1983,ugedaCharacterizationCollectiveGround2016}, 
and the $d_{2,0}$ character of the orbitals at the $\Gamma$ valley. 
Despite the quasiparticle spectrum being fully gapped, the strong variation of the order parameter on the $\Gamma$ Fermi surface leads to tunneling spectra mimicking a nodal gap, i.e., 
 a $V$-like shape of the $\T{d}I/\T{d}V$ as observed in the experiment. Indeed, the same spectroscopy data could be fitted by assuming a standard nodal $f$-gap~\cite{vanoEvidenceNodalSuperconductivity2023}. 
 By using the nematic solution or the subleading $s+f$ gap function, a similar good fit can be  obtained as well, as seen in Fig.~\ref{fig: STS_Ta}(c). 
 
   \textit{Conclusions}---Our results show that a mechanism purely based  on long-range screened Coulomb repulsion can reconcile theoretical expectations of  chiral superconductivity in 2D triangular lattices with seemingly contradicting low-temperature STS experiments. To this extent, it is crucial to account for the Ising SOC present in the monolayer TMDCs, which mixes nominally even and odd solutions and yields a dispersive, albeit gapped, chiral phase.
Despite the above results being promising, they are not yet conclusive:  
further experiments  are needed to confirm the chiral nature of the gap, while a full theoretical treatment should include electron- and phonon-mediated mechanisms on the same footing. 

While the focus of this work was on intrinsic TMDCs superconductors, the method developed here is more general and could be easily applied to other low-dimensional superconductors. A natural candidate is monolayer  MoS$_2$, a direct gap semiconductor which becomes superconducting upon doping~\cite{luEvidenceTwodimensionalIsing2015} and for which a Kohn-Luttinger mechanism 
has been suggested ~\cite{roldanInteractionsSuperconductivityHeavily2013}. Furthermore, evidence for chiral $d$-wave superconductivity has been reported for tin monolayers on a Si(111) surface~\cite{mingEvidenceChiralSuperconductivity2023}, and a transition from a nematic to a chiral phase has been proposed for the layered 4Hb-TaS$_2$~\cite{ribakChiralSuperconductivityAlternate2020,silberTwocomponentNematicSuperconductivity2024}. 
\\

{\textit{Acknowledgments}}---Funding was provided by the CRC 1277 and RTG 2905 - 502572516 of the Deutsche-Forschungsgemeinschaft. 
The experiments made use of the Aalto Nanomicroscopy Center (Aalto NMC) facilities with support from the Academy of Finland (Academy research fellow No. 361420). \\

{\textit{Data Availability}}---The data that support the findings of
this work  will be made available
 upon request to the corresponding author Milena Grifoni. 
%
%
%

%
\onecolumngrid
\begin{center} 
\subsection*{End Matter} 
\end{center} 
\twocolumngrid
%
%
{\textit{Parametrization of the LCAO tight-binding model}}---We present details regarding the multi-orbital tight-binding calculation of the relevant low-energy bands of 1H-NbSe$_2$ and 1H-TaS$_2$ shown in Fig.~\ref{fig:bands}.    For  the former, the bands are obtained from a linear combination of the three atomic $d$-orbitals of Nb  $4d_{z^2}=4d_{2,0}$ and $4d_{2,\pm 2} = 4d_{x^2-y^2} \pm i 4d_{xy}$. In TaS$_2$ the orbitals come from the next shell, and hence we consider $5d_{z^2}=5d_{2,0}$ and $5d_{2,\pm 2} = 5d_{x^2-y^2} \pm i 5d_{xy}$. For the functional form of the orbitals we use the one of the hydrogen atom with  effective nuclear charge $Z_{\rm eff}$ from~\cite{clementiAtomicScreeningConstants1967}.  
The tight-binding parametrization by \citeauthor{heMagneticFieldDriven2018}~\cite{heMagneticFieldDriven2018} is adopted for 1H-NbSe$_2$, but similar results are obtained from the one reported by~\citeauthor{kimQuasiparticleEnergyBands2017}~\cite{kimQuasiparticleEnergyBands2017}. For 1H-TaS$_2$ we rely on the results by~\cite{margalitTheoryMultiorbitalTopological2021}. \\

\renewcommand{\theequation}{B\arabic{equation}}
\setcounter{equation}{0} 
\textit{Matrix elements of the screened interaction}---In the Bloch basis the matrix elements of the screened interaction have the form 
 \begin{align}
    \label{matrix element bloch states expanded}
    &\bra{\Qm(\bk+\bq),\sigma;\Qm(\bk'-\bq),\sigma'}\hat{V}\ket{\bk,\sigma;\bk',\sigma'}=\\
    &\frac{1}{N\Omega}\sum_{\bG,\bG'}V_{\bm{G},\bm{G}'}^{\T{2D,RPA}}(\bq;0^+)\Fm^{\sigma}_{\bk+\bq,\bk}(-\bm{G})\Fm^{\sigma'}_{\bk'-\bq,\bk'}(\bm{G}')\,,\nonumber
\end{align}
where $\bk,\bk'$ and $\bq$ are in-plane momenta restricted to the first Brillouin zone.
Furthermore, the projector $\Qm$ ensures that the scattered momenta are folded back into the first Brillouin zone as well.  We omit to indicate the projector $\cal{Q}$ in the following for simplicity. 
 The Bloch overlaps $\Fm$  are defined as
\begin{align}
    \Fm^{\sigma}_{\bk,\bk'}(\bm{G}) =&\int_{\cal{V}_p}d\bm{r}e^{-i\bm{G}\cdot\bm{r}}\bm{u}^\dagger_{\sigma,\bk}(\bm{r})\bm{u}_{\sigma,\bk'}(\bm{r})\,,
    \label{overlap factors}
\end{align}
with $\bm{u}_{\sigma,\bk}$ Bloch spinors,  and the unit cell volume ${\cal{V}_p}$.

The matrix elements in~\cref{matrix element bloch states expanded} 
are in general complex, with a strongly fluctuating phase resulting from the product of the Bloch overlaps.
This tends to favor pairing between time-reversed states, whose Bloch spinors obey the relation $\bm{u}_{\bar{\sigma},\bar{\bk}} = \bm{u}_{\sigma,{\bk}}^*$, with  $-\bm{k}=:\bar{\bm{k}},-\sigma=:\bar{\sigma}$,  resulting in a summation of real and positive contributions.  
We hence retain only the scattering between time-reversal partners ($\bm{k},\sigma;\bar{\bm{k}}, \bar{\sigma}$), and consider the following interaction matrix elements 
${V_{\bk,\bk',\sigma}:=\bra{\bk',\sigma';\bar{\bk}',\bar\sigma'}{\hat{V}}\ket{\bk,\sigma;\bar{\bk},\bar{\sigma}}}$. 
\\

\textit{Experimental setup and methods}---TaS$_2$ monolayers$_2$ were grown on highly oriented pyrolytic graphite (HOPG)
using molecular beam epitaxy (MBE). After growth, the sample was inserted into
the low-temperature STM (Unisoku USM-1300) housed in the same ultra-high vacuum 
system and subsequent experiments were performed at $T$ = 350 mK.
The $dI/dV$ spectra were
recorded by standard lock-in detection while sweeping the sample bias in
an open feedback loop configuration, with a peak-to-peak bias modulation
specified for each measurement and at a frequency of 911 Hz. The $dI/dV$ 
spectra were recorded on extended monolayers to eliminate possible island
size dependence from the measurements. 
The data shown in Fig.~\ref{fig: STS_Ta} are normalized by the averaged $dI/dV$. 
\\

\renewcommand{\theequation}{D\arabic{equation}}
\setcounter{equation}{0} 
\textit{Comparison to experimental STS data}---For tunneling spectroscopy with a normal metal tip, the differential conductance is well approximated  by
\begin{align}
    \label{fit function}
    G(V)&\approx\sum_{\pi}C_{\pi}G_\pi(V)\,,\\
    G_\pi(V)&=\int_{-\infty}^{\infty}\T{d}E\;D_{\pi}(E)\left(-\frac{\partial f(E+eV)}{\partial E}\right)\,,
    \label{differential conductance per transport channel channel}
\end{align}
where 
the couplings $C_{\pi}$, account for the spectral function of the tip and the tunneling overlap between the tip's evanescent states and quasiparticle states from the superconductor on a given Fermi surface $\pi=K,\Gamma$.
The tunneling to each of the Fermi surfaces represents a distinct transport channel, whose strength is given by the sum over the local quasiparticle  density of states
\begin{align}
    \label{angleDependentDOS}
    D_{\pi}(E)=\sum_{\bk\in\pi,\sigma}D_{\bk,\sigma}(E)\,.
\end{align}
The latter are modeled by assuming a BCS form factor modifying the local density of states $\rho_{\bk,\sigma}$ in the normal conducting state. 
In order to reproduce the experimentally measured critical temperature of $T_c=1\,$~K 
~\cite{
vanoEvidenceNodalSuperconductivity2023}, we used a gap obtained by 
 solving the gap equation with the interaction potential $\hat{V}$ multiplied by a factor $\gamma$, which accounts, for example, for  charge transfer from the substrate  or an additional phonon-mediated interaction~\cite{hinscheElectronPhononInteraction2017,dasElectronphononCouplingSpin2023}, which have not been explicitly included in our calculation.

Due to the different orbital composition of the quasiparticle states at the  $K$, $K'$ valleys and at the $\Gamma$ surface~\cite{liuThreebandTightbindingModel2013}, the coupling constant will be in general different.
In particular, since the $d_{2,0}$ orbital extends farthest out of plane and the STM favors states with small in-plane momentum~\cite{ugedaCharacterizationCollectiveGround2016,tersoffTheoryApplicationScanning1983}, we expect $C_{\Gamma}\gg C_{K}$. 
For low temperature, the form of the gap on each separate Fermi surface results in characteristic signatures in the differential conductance of the respective transport channel.
We use as a fit function this superposition of the differential conductance of the individual transport channels. 
For the gaps on the Fermi surfaces $\Delta_{\bk,\sigma}$, we use the form of the gaps as found from \cref{gap equation trs} at the experimental temperature $T_{\T{exp}}$
and allow for a fit of the amplitude by a common rescaling of all $\Delta_{\bk,\sigma}$ by a parameter $A$ as $\Delta_{\bk,\sigma}\rightarrow A\Delta_{\bk,\sigma}$. 
The latter is introduced to take care of the uncertainty of the actual experimentally realized $T_{\T{c}}$ and the rescaling factor $\gamma$ of the interaction.

To qualitatively account for additional sources of broadening in the experiment, we fit with an effective temperature $T_{\T{eff}}$ in the calculation of the derivative in~\cref{differential conductance per transport channel channel}, about twice  the recorded base temperature of the STM. 
To account for offsets in the calibration, we further allow for both a small constant offset $G_0$ in the measured conductivity and $V_0=0.009$~mV in the recorded voltage. 
For the fitting we use the trust region reflective algorithm as implemented in SciPy's "curve\_fit" routine.
We consider  a range of $\pm 0.5\,\T{mV}$ containing the main coherence peaks. 
The resulting fit parameters are listed in~\cref{table fit coefficients}.
The best fits for all the solutions  are obtained by considering strongly selective coupling of the tip to the $\Gamma$ pockets.\\

\begin{table}[h]
\centering
\begin{tabular}{|l|c|c|c|c|c|}
\hline
Gap & $C_K$ , $C_\Gamma$ (eV \AA$^2$) & $A$  & $G_0$ & $T_{\rm eff}/T $&  $\gamma $ \\
\hline
chiral  & 1.607, 15.041 & 1.110 & 0.173 & 2.175  & 9.298 \\
\hline
nematic & 0.000, 17.235 & 1.156  & 0.174 & 1.815  & 9.298 \\
\hline
$s+f$  & 1.890, 18.333 & 1.017  & 0.000 & 1.804  & 8.777 \\
\hline
\end{tabular}
 \caption{
    Coefficients of best fit between the theoretical prediction of the differential conductance in the possible superconducting phases and the experiment as shown in~\cref{fig: STS_Ta}\,(b), (c).
    The best fit for each solution is achieved by selective coupling to the $\Gamma$ pocket.
    }
    \label{table fit coefficients}
\end{table}

\end{document}